\documentclass[12pt]{article}
\usepackage{graphicx}
\begin{document}
version 22.1.2014
\begin{center}
{\Large Lepton bound states in a fundamental \\description
of elementary forces }   
\vspace{0.3cm}

H.P. Morsch \\ HOFF, Brockm\"ullerstr.~11,
D-52428 J\"ulich, Germany\\  E-mail:
h.p.morsch@gmx.de 
\end{center}

\begin{abstract}
In the flavour dependence hadrons and leptons show a mass relation,
not expected in the Standard Model. 
This can be understood in an analysis, based on a particle
Lagrangian with Maxwell term, boson-boson coupling and massless fermions
(quantons), in which hadrons are described as stationary systems of quantons
bound by an electric interaction, whereas leptons represent systems bound
magnetically. 

PACS/ keywords: 11.15.-q, 12.10.-g, 14.60.-z/ Hadron-lepton relation 
in the flavour dependence, well understood in a theory based on
a Lagrangian with Maxwell term, boson-boson coupling and massless elementary
fermions (quantons). Composite systems bound by electric and magnetic forces. 
\end{abstract}

During the last decades our knowledge of the features of fundamental forces
has increased largely due to strongly improved detector technologies, 
accelerators with higher energetic beams and better telescopes and new
technologies in astrophysics. However, in the theoretical understanding the
Standard Model of particle physics~\cite{PDG} (SM), established 40 years ago,
still represents the state of our knowledge. It is a heuristic model
constructed from first order gauge theories, different for each fundamental
force (but gravitation could not be integrated in the SM). Starting from
quantum electrodynamics (QED), each additional force requires a different
Lagrangian with additional fields. Further, defaults in these theories to
describe the flavour degree of freedom and fermion masses require still more
fields. Mass has been explained by the Higgs-mechanism, but neutrino masses
need also the postulation of heavy Majorana neutrinos. In total, this indicates
clearly that the SM (which has many parameters, which cannot be 
determined within the model) is far from a fundamental theory of elementary
forces, which should have no free parameters. In addition, a fundamental
theory should have very few elementary fields, since close to the origin
nature is expected to develop only from fields, which are present in the
vacuum or can be generated out of it. 
Experimental evidence for the additional fields needed in the SM have not been
found\footnote{Concerning evidence for the Higgs-boson see ref.~\cite{Mo3}.}.  

Even QED, the best known part of the SM, cannot be regarded as a
fundamental theory, since the coupling constant $\alpha_{QED}\sim$1/137 is not
understood from first principles. However, the precise prediction of spin
properties suggests that QED is close to a fundamental theory. Only
the divergencies at $r\to 0$ and $\infty$ appear to be in conflict with
nature, which is known to develop in a smooth and finite way. Because of this,
a fundamental theory is expected to have a more complex structure
than the first order gauge theories in the SM.

A severe problem in the SM is the non-symmetry of hadrons and leptons. 
Hadrons are assumed to be complex particles composed of
elementary fermions (quarks) bound by the strong interaction,
whereas leptons are themselves considered as elementary fermions, 
which couple only by the weak interaction. However, in the flavour dependence
of hadron and lepton masses in fig.~1 a relation may be seen (for increasing
flavour number hadron and charged lepton masses approach each other), which
could reveal that these particles are not completely independent. This is
expected, since on a rather fundamental level hadrons and leptons should have
a similar origin.

Quite recently a second order extension of QED has been
studied~\cite{Mo1,Mo2}, which has been found to show all features expected of
the long-sought fundamental theory. It is finite and capable to describe 
systems bound by different forces. Further, with massless elementary bosons
and fermions (quantons) this model shows a coupling to the vacuum
with average boson and fermion energies $E_{vac}=0$, and very important, it has
no free parameters. Applied to the binding of light atoms~\cite{Mo4}, the
self-consistently deduced coupling constant $\alpha$ is consistent with
$\alpha_{QED}$, thus leading to an understanding of the fine structure constant.
The flavour degree is contained in a discrete ambiguity of the
radial extent of bound state solutions, the sum of which is constrained by a
vacuum potential sum rule, see ref.~\cite{Mo4}. 

It has to be mentioned that in the past different Lagrangians have been
studied with the conclusion that higher order theories should be generally
discarded, because they lead to unphysical solutions~\cite{highd}. However,
the important difference to the higher order theories discussed in
ref.~\cite{highd} is that in the present formalism the Lagrangian is gauge
invariant and non-physical solutions can
be eliminated by strict geometric, mass-radius and energy-momentum relations. 

The Lagrangian has been used in the form 
\begin{equation}
\label{eq:Lagra}
{\cal L}=\frac{1}{\tilde m^{2}} \bar \Psi\ i\gamma_{\mu}D^{\mu}
D_{\nu}D^{\nu}\Psi\ -\ \frac{1}{4} F_{\mu\nu}F^{\mu\nu}~,   
\end{equation}
where $\tilde m$ is a mass parameter and $\Psi$ a two-component massless
fermion field, $\Psi=(\Psi^+, \Psi^o)$ and $\bar \Psi= (\Psi^-, \bar \Psi^o)$,
with charge and neutral part. 
Vector boson fields $A_\mu$ with coupling $g$ to fermions are contained in the
covariant derivatives $D_{\mu}=\partial_{\mu}-i{g} A_{\mu}$. 
The second term of the Lagrangian represents the Maxwell term with
Abelian field strength tensors $F^{\mu\nu}$ given by $F^{\mu\nu}= 
\partial^{\mu}A^{\nu}-\partial^{\nu}A^{\mu}$, which gives rise to both electic 
and magnetic effects. 

By inserting $D^{\mu}=\partial^{\mu}-i{g} A^{\mu}$ and
$D_{\nu}D^{\nu}=\partial_{\nu}\partial^{\nu}
-ig(A_{\nu}\partial^\nu+\partial_\nu A^\nu) -g^2 A_\nu A^\nu$ in
eq.~(\ref{eq:Lagra}), the first term of ${\cal L}$ gives rise to a number of
different terms, which contain boson and fermion fields and/or their
derivatives, see ref.~\cite{Mo2}. All terms, which contain the
derivative of the fermion field $\partial^{\nu} \Psi$, are related to a rather
complex dynamics of the system. For stationary solutions, of only interest
here, only two terms of the Lagrangian contribute  
\begin{equation}
\label{eq:L2}
{\cal L}_{2g} =\frac{-ig^{2}}{\tilde m^{2}} \ \bar \Psi \gamma_{\mu}
[A^\mu \partial_{\nu} A^\nu]~\Psi \ 
\end{equation}
and 
\begin{equation}
\label{eq:L3}
{\cal L}_{3g} =\frac{ -g^{3}\ }{\tilde m^{2}} \ \bar \Psi \gamma_{\mu}
[A^\mu A_\nu A^\nu]~\Psi \ .
\end{equation}
The gauge condition $\partial_\mu A^\mu=0$ used for first order Lagrangians
is replaced in the present case by $\partial(\partial_\nu A^\nu)=0$. 

Because of its high complexity a general study of the properties of the
Lagrangian~(\ref{eq:Lagra}) appears to be rather difficult. So far only 
fermion matrix elements from the Lagrangians~(2) and (3) have been evaluated
(derived from generalised Feynman diagrams, see e.g.~ref.~\cite{matrix}). 
However, this has been found to be an efficient and reliable method
to generate bound state potentials from the Lagrangian. These matrix elements
have been used in the  
form ${\cal M}=<g.s.|~K(\tilde p'-\tilde p)~|g.s.>\sim \bar 
\psi(\tilde p')\ K(\tilde p'-\tilde p)~\psi(\tilde p)$, where $\psi(\tilde p)$
is a fermion wave function $\psi(\tilde p)=\frac{1}{\tilde m^{3/2}}
\Psi(p)\Psi(k)$ and $K(q)$ a kernel given by  
$K(q)=\frac{1}{\tilde m^{2(n+1)}}\ [O^n(q)\ O^n(q)]$, were n is the number of
derivatives and/or boson fields. For the above Lagrangians~(\ref{eq:L2}) and
(\ref{eq:L3}) n=3 and $O^3(q)=\gamma_{\mu} A^\mu \partial_{\nu} A^\nu$
and $\gamma_{\mu} A^\mu A_\nu A^\nu$, respectively. This leads to 
\begin{equation}
\label{eq:M2}
{\cal M}_{2g} =\frac{\alpha^{2}}{\tilde m^8} \bar \psi(p')\gamma_\mu
A^{\mu}(q)~(\partial_\nu A^{\nu}(q))(\partial_\sigma
A^{\sigma}(q))~\gamma_\rho A^{\rho}(q) \psi(p)\ 
\end{equation}
and
\begin{equation}
\label{eq:M3}
{\cal M}_{3g} = \frac{-{\alpha}^{3}}{\tilde m^8}~\bar
\psi(p')\gamma_\mu A^\mu(q)~A_\nu(q)A^\nu(q)~
A_\sigma(q)A^\sigma(q)~\gamma_\rho A^\rho(q) \psi(p) \ , 
\end{equation}
where $\alpha=g^2/4\pi$. 

Before these matrix elements are discussed in detail, one may examine 
the properties of a simpler (first order) Lagrangian of the form
${\cal L}_{f.o.}= \bar \Psi\ i\gamma_{\mu}D^{\mu}
\Psi\ -\ \frac{1}{4} F_{\mu\nu}F^{\mu\nu}$. This Lagrangian leads
only to one matrix element ${\cal M}_{f.o.} =
\frac{-{\alpha}}{\tilde m^4}~\bar 
\psi(p')\gamma_\mu A^\mu(q)~\gamma_\rho A^\rho(q) \psi(p)$. By adding a matrix
element with interchanged $\mu$ and $\rho$ (using 
$\frac{1}{2} (\gamma_\mu\gamma_\rho+ \gamma_\rho\gamma_\mu) =g_{\mu\rho}$) the
$\gamma$-matrices can be removed. Further, by an equal
time requirement of the two boson fields\footnote{in a
  ($t,\vec r$) representation} the matrix element can be written in the form
${\cal M}^f_{fo} = \bar \psi(p')~V_v(q)~\psi(p)$, where $V_v(q)$ can be
interpreted as boson-exchange interaction of vector structure
$V_v(q)=\frac{-{\alpha}}{\tilde m^4} A_\mu(q)A^\rho(q)$. Since the two boson
fields are relativistic, they overlap only momentarily and should not form 
a stable bound state potential $V_v(q)$. Therefore, the existence of a
boson-exchange potential in form of the Coulomb potential $V_{coul}(r)=
\alpha (\hbar c)~1/r$ between two relativistic particles with distance $r$
is not well understood (a non-relativistic approximation for the
interaction of relativistic particles, like electrons and protons, is not
justified a priori). 
If the charged particles are in motion with a velocity v relative to the
potential, also magnetic effects occur. This leads to the Ampere force
law $F_m(r)=\frac{\mu_o (\hbar c)}{2\pi} I^2/r$. Using $I=q\frac{v}{c}c$ and 
the Maxwell relation $\epsilon_o\mu_o c^2=1$, the Ampere potential
$V_m(r)=\alpha (\hbar c)~(\frac{v}{c})^2/r$ is obtained, which has a
structure similar to the Coulomb potential, but multiplied with a factor
$(\frac{v}{c})^2$. Stationary particle states bound in such a potential have
never been seriously considered.

Now possible bound state potentials for the second order
Lagrangian~(\ref{eq:Lagra}) are discussed by inspecting the structure of the
matrix elements ${\cal M}_{2g}$ and ${\cal M}_{3g}$. Again the
$\gamma$-matrices can be removed by adding a matrix element with interchanged
$\mu$ and $\rho$. Further, the derivatives of boson fields in ${\cal M}_{2g}$
may be written in the form $(\partial_\nu A^{\nu}(q))(\partial_\sigma A^{\sigma}(q))
=\frac{1}{2} \partial_\nu [\partial_\sigma (A_\mu A^\mu)^\sigma]^\nu$, using
the above gauge condition. In addition, the two boson fields on the right and
left of eq.~(\ref{eq:M3}) can be combined (analogue to the fermion wave
functions) to (quasi) wave functions\footnote{with
  dimension $[GeV]$. Further, for bosons $\bar W_{\mu}^\nu(q)=W_{\mu}^\nu(q)$.}
$W_{\mu}^\nu(q)= \frac{1}{\tilde m} A_\mu(q) A^\nu(q)$, whereas the remaining
two boson fields may be understood as a boson-exchange interaction of vector 
structure $V_{\mu}^\nu(q)\sim W_{\mu}^\nu(q)$ ($\mu\neq\nu$) similar to that
of first order theory discussed above. The fact that two
boson fields can be combined to wave functions leads quite naturally to a
finite theory, in which the wave functions are normalised. 
 
Similar to the first order case, non-vanishing matrix elements can be obtained
only, if there is spacial overlap of the boson fields at a given time.
However, the important difference to the first order case is that in addition
to the boson-exchange matrix element ${\cal M}_{3g}$ a second term of
derivative structure ${\cal M}_{2g}$
exists, which leads to a dynamical stabilisation of the system. If a $q\bar q$
pair is created momentarily, this term leads to confinement, since the
corresponding binding energy is positive. Consequently, the created fermion
pair is locked in a bound state. 
The equal time requirement gives rise to a reduction of the fermion
four-vectors to three-vectors in momentum or r-space, while the boson vectors
are reduced to two dimensions. The boson wave
functions $W_{\mu}^\nu(q)$ give scalar and vector components $w_{s}(q)$ and
$w_{v}(q)$, whereas $V_{\mu}^\nu(q)$ yields an interaction potential
$v_{v}(q)\sim w_{v}(q)$. In this way, the fermion 
matrix elements~(\ref{eq:M2}) and (\ref{eq:M3}) can be written by 
\begin{equation}
\label{eq:P2g}
{\cal M}^f_{2g} =\frac{\alpha^{2}}{2\tilde m^{6}} \bar \psi(p')
\ w_{s,v}(q) \partial^2 w_{s,v}(q)\ \psi(p)\ ,
\end{equation}
and
\begin{equation}
\label{eq:M3f}
{\cal M}^f_{3g} = \frac{-{\alpha}^{3}}{\tilde m^5}~\bar
\psi(p')w_{s,v}(q)~v_{v}(q)~w_{s,v}(q) \psi(p) \ .
\end{equation}
The bosonic part of eq.~(\ref{eq:M3f}) can also be written in the form
of a matrix element, in which the wave functions $w_{s,v}(q)$ are
connected by $v_v(q)$
\begin{equation}
\label{eq:P2g'}
{\cal M}^{g} =\frac{-\alpha^{3}}{\tilde m^{2}}
\ w_{s,v}(q) v_{v}(q) w_{s,v}(q) .
\end{equation}
This matrix element shows binding of two bosons in the potential $v_{v}(q)$.
According to the virial theorem the term 
$\partial^2 w_s(q)$ in eq.~(\ref{eq:P2g}) is related to the kinetic energy of
these bosons. 
In a transformation to r-space the bosonic part of eq.~(\ref{eq:P2g}) gives
rise to a Hamiltonian, which may be written in the form
\begin{equation}
\label{eq:H}
-\frac{\alpha^2 (\hbar c)^2 F_{2g}}{4\tilde m}~\Big (\frac{d^2
    w_s(r)}{dr^2} + \frac{2}{r}\frac{d w_s(r)}{dr}\Big ) +V_{2g}(r)~w_s(r)
  = E_i~w_s(r)~,
\end{equation}
where the factor $F_{2g}$ is due to the Fourier transformation of the boson
kinetic energy. This leads to a binding potential
\begin{equation}
V_{2g}(r)= \frac{\alpha^2 (\hbar c)^2 F_{2g}}{4\tilde m}\ \Big
(\frac{d^2 w_s(r)}{dr^2} + 
  \frac{2}{r}\frac{d w_s(r)}{dr}\Big )\frac{1}{\ w_s(r)}+E_o\ , 
\label{eq:vb}
\end{equation}
where $E_o$ the energy of the lowest eigenstate. A connection to the vacuum
can be made by assuming $E_o=\tilde E_{vac} = 0$.  
This potential leads to a stabilisation of the system. It can be identified
with the confinement potential, required in hadron potential models~\cite{qq}
and shows an almost linear increase towards larger radii.

Writing the matrix element ${\cal M}^f_{3g}$ in the form ${\cal M}^f_{3g} = \bar
\psi(p')\ V_{3g}(q)\ \psi(p)$, this leads to a three-boson potential
\begin{equation}
\label{eq:V3}
{ V}^{s,v}_{3g}(q)=\frac{-\alpha^3}{\tilde m^{2}}~w_{s,v}^2(q) v_v(q)\ 
\end{equation}
with the interaction $v_v(q)\sim w_v(q)$. Fourier transformation to r-space
yields a folding potential   
\begin{equation} 
\label{eq:vqq}
V^{s,v}_{3g}(r)= -\frac{\alpha^3 \hbar c}{\tilde m} \int dr'
w_{s,v}^2(r')\ v_v(r-r')~.   
\end{equation}

From the general structure of the fermion matrix element in eq.~(\ref{eq:M3f}) 
one can see that there are two $q\bar q$ states (with quantum
numbers $J^\pi=1^-$) with scalar and vector boson wave functions
$w_{s,v}(r)$ and corresponding fermion wave functions\footnote{for the radial
  wave functions $\bar \psi(r) =\psi(r)$.} $\psi_{s,v}(r) \sim w_{s,v}(r)$. 
Further, there are two $q\bar q$ p-states (with quantum numbers $J^\pi=0^+$)
with similar wave functions, see e.g.~ref.~\cite{Mo3}.

The fermion wave functions have to be orthogonal, leading to the constraint
\begin{equation}
\label{eq:ortho}
\int r^2dr~\psi_s(r) \psi_v(r)=\int r^2dr~w_s(r) w_v(r)=<r_{w_s,w_v}>=0 \ .
\end{equation}
To satisfy this condition, $w_{v}(r)$ may be written in the form of a
p-wave function 
\begin{equation}
\label{eq:spur}
w_{v}(r) = w_{v,o}~[w_s(r)+\beta R\ \frac{d w_s(r)}{dr}]~,
\end{equation}
where $w_{v,o}$ is obtained from the normalisation $2\pi \int r dr\ w_v^2(r)
=1$ and $\beta R$ determined by $\beta R=-\int r^2dr~w_s(r)/\int
r^2dr~[dw_s(r)/dr]$. 
Interestingly, orthogonality gives rise to another quite natural condition for
the deepest bound state, requiring that the interaction takes place inside the
bound state volume of $w_s^2(r)$. This leads to 
\begin{equation}
\label{eq:conr}
|V^v_{3g}(r)| \sim {c}\ w^2_s(r)  \ . 
\end{equation}

The conditions~(\ref{eq:ortho})-(\ref{eq:conr}) lead to a
boson wave function $w_s(r)$, which can be approximated by 
\begin{equation}
\label{eq:wf}
w_s(r)=w_{s,o}\ exp\{-(r/b)^{\kappa}\} \ , 
\end{equation} 
where $w_{s,o}$ is fixed by the normalisation $2\pi \int
r dr\ w_s^2(r) =1$. The parameters $b$ and $\kappa$ have to be determined by
boundary conditions as discussed below. Different flavour states are
obtained by solutions with different slope parameter $b$, which are
constrained by a vacuum potential sum rule, see e.g.~ref.~\cite{Mo4}.
The interaction $v_v(r)$ is given by $v_v(r)=\hbar c~w_v(r)$.

Binding energies have been calculated from the potentials $V_{ng}(r)$ by using
the virial theorem in the form $4\pi[\int r^2dr~\psi^2(r)V_{ng}(r)
  -\frac{1}{2}\int r^3dr~\psi^2(r)\frac{d}{dr}V_{ng}(r)]=E_f$, where $\psi(r)$
are fermion wave functions with a form similar to the boson wave
functions in eqs.~(\ref{eq:spur}) and (\ref{eq:wf}). In addition,
eq.~(\ref{eq:P2g'}) shows that 
$V^{s}_{3g}(r)$ can be interpreted as bound state of bosons. Its binding
energy $E_g$ has been calculated by the corresponding form 
$2\pi[\int rdr~w_s^2(r)v_v(r) -\frac{1}{2}\int
  r^2dr~w_s^2(r)\frac{d}{dr}v_v(r)] =E_g$.
For massless fermions the mass
of the system $M^{s,v}$ is given by the absolute binding energies in
$V_{2g}(r)$ and $V^{s,v}_{3g}(r)$, $M^{s,v}=-E^{3g}_{f_{s,v}}+E^{2g}_f$, while the
reduced mass is given by $\tilde m=M^{s}/2$. 

In order to make an evaluation of the potentials other constraints are needed,
which connect the coupling constant $\alpha$ to the shape parameters and the
mass parameter $\tilde m$. $(q\bar q)^n$ systems can be bound by an
electric interaction\footnote{with an equivalent coupling constant $\alpha_{eq}
  \geq$1/137.}. One condition is the energy-momentum relation, important for  
relativistic bound states. For binding in the potential $V^s_{3g}(q)$ the
negative fermion and boson binding energies $E_f^s$ and $E_g$ have to be
compensated by their root mean square momenta $<q^2_{v}>=\int dq~q^3~v(q)/\int
dq~q~v(q)$ in the potentials $V_{3g}(r)$ and $v_{v}(r)$, respectively
\begin{equation}
<q^2_{V_{3g}}>+<q^2_{v_v}> = (E_f^s+E_g)^2 \ . 
\label{eq:massq}
\end{equation}
Another condition can be derived from the confinement potential~(\ref{eq:vb}),
see ref.~\cite{Mo2}, which leads to 
\begin{equation}
Rat_{conf}=\frac{(\hbar c)^2}{\tilde m^2 <r^2_{w_s}>} =1 \ .
\label{eq:ravb}
\end{equation}
By these conditions all parameters are interrelated and can be uniquely
determined for a given mass of the system. The above formalism has
been discussed for electric binding of quantons in hadronic
$q\bar q$ systems in ref.~\cite{Mo2,Mo3}, further for the description of light
atomic systems in ref.~\cite{Mo4}. Importantly, for atoms the deduced coupling
constant has been found to be consistent with $\alpha_{QED}\sim 1/137$, thus
leading to an understanding of this constant from first principles. For the 
much smaller hadronic $q\bar q$ bound states the magnitude of the coupling
constant is significantly larger. 

Now a new situation is discussed, in which the fermion fields are in relative
motion with velocity v to lead to stationary systems, which
are bound magnetically. This is not possible for uniform motion of the total
density, e.g.~as a rotation of a $q\bar q$-system (possible for an
electrically bound system), but requires a more complex system with at least
two fermions, in which the positive and negative fermions move in opposite
direction to each other with a relative velocity $({v}/{c})$. This is possible
in $(q\bar q)^n q$ systems, requiring that the fermion momenta add up to zero,
$\sum_{i=1,n}\vec{<q_{\psi_i}>}=0$. Described by fermion densities 
this leads to $<q^2_{V_{3g}}>=0$. Due to the opposite motion of positive and
negative fermions, on the average all electric interactions cancel out. 
With a reduction of the
potentials by a factor $({v}/{c})^2$ (see the first order case discussed
above) this gives rise to an energy-momentum relation
\begin{equation}
<q^2_{v_{v}}> ({v}/{c})^2 = (E_f^s+E_g)^2 \ . 
\label{eq:mqmag}
\end{equation}
Another difference from electric bound states, p-wave states with nodes in the
wave functions do not lead to stable magnetic bound states, thus allowing only
{\bf one} stable magnetic state with boson and fermion wave functions $w_s(r)$
and $\psi_s(r)$. Further, the confinement potential leads to the condition
\begin{equation}
Rat_{conf}=\frac{(\hbar c)^2 ({v}/{c})^2}{\tilde m^2 <r^2_{w_s}>} =1 \ .
\label{eq:ratmag}
\end{equation}
Finally, $({v}/{c})^2$ has to be included in the bound state potential 
$V_{3g}(r)$. Altogether this gives three conditions, by which $({v}/{c})^2$ is
determined 
\begin{equation}
({v}/{c})^2_{conf} = \frac{\tilde m^2 <r_{w_s}^2>}{(\hbar c)^2} \ ,
\label{eq:mag1}
\end{equation}
\begin{equation}
({v}/{c})^2_{mom} = (E^s_f+E_g)^2/<q^2_{v_{v}}> \ ,
\label{eq:mag2}
\end{equation}
\begin{equation}
({v}/{c})^2_{pot} = (\alpha_{fit}^3)/\alpha^3 \ ,
\label{eq:mag3}
\end{equation}
where $\alpha_{fit}^3=\alpha^3({v}/{c})^2$ is the coupling constant in
eq.~(\ref{eq:vqq}) adjusted to get the binding 
energy $E_f^s$. Only if the same value of $({v}/{c})^2$ is obtained in all
three expressions~(\ref{eq:mag1})-(\ref{eq:mag3}), a stable bound
state is created.

The above formalism has been used to describe the structure of charged and 
neutral leptons $e$, $\mu$, $\tau$ and $\nu_e$, $\nu_\mu$, $\nu_\tau$, 
which are assumed as stationary $(q\bar q)^n q$ systems 
bound magnetically. The shape parameter $\kappa$ has been taken from the
analysis of hadrons~\cite{Mo2}, whereas the slope parameter and the
coupling constant $\alpha$ has been adjusted to give the same value of 
$({v}/{c})^2$ from all three conditions~(\ref{eq:mag1})-(\ref{eq:mag3}). 

First, neutrino bound states are discussed. Starting from a pure $q^o\bar q^o
q^o$ structure, magnetic binding without motion of charge is not
possible. However, a $q^o\bar q^o$ pair can change to two $(q^+ q^-)$ pairs,
in which the positive and negative fermions move in opposite direction to each
other with relative velocity $v/c$. This can lead to magnetic binding. 
Indeed, in such an analysis the three boundary 
conditions~(\ref{eq:mag1})-(\ref{eq:mag3}) can be fulfilled (which is far from
trivial), leading to a system with a very small radius. This is conceivable,
since the magnetic force is only strong enough at extremely small
distances. First, the slope parameter $b$ has been varied to get
the same value of $({v}/{c})^2$ from the relations~(\ref{eq:mag1}) and
(\ref{eq:mag2}); then $\alpha$ has been adjusted to get the same value of
$({v}/{c})^2$ from relation~(\ref{eq:mag3}). This was possible for all three
neutrinos and confirms the conjecture that these 
systems are bound magnetically. The used parameters, masses, deduced
mean square radii and values of $({v}/{c})^2$ are given in table~1.
The masses have been used from ref.~\cite{Moneu}, which yield
mass square differences $m_{\nu_\mu}^2-m_{\nu_e}^2=\Delta m^2_{sol}$ and
$m_{\nu_\tau}^2-m_{\nu_e}^2= \Delta m^2_{atm}$ consistent with the values
$\Delta m^2_{sol}\simeq 7.6\ 10^{-5} (eV)^2 $ and $\Delta m^2_{atm}\simeq
2.4\ 10^{-3} (eV)^2$ deduced from neutrino oscillation
experiments~\cite{neuex}. 

Results on the shape of the interaction $v_v(r)$ is given in the upper 
part of fig.~2 in comparison to the $1/r$ dependence of the Ampere
potential. In the middle part the r-dependence of boson density and potentials
$V_{eg}(r)$ are displayed, which shows that relation~(\ref{eq:conr}) is well
fulfilled. The confinement potential $V_{2g}(r)$ is shown in the 
lower part, which has a form rather similar to that deduced for hadrons.
Although the absolute magnitude is rather low for very small bound states,
the positive binding energy in this potential is responsible for stabilisation
of the system.  
\begin{table}
\caption{Parameters ($b$ in fm) and masses, mean square radii (in fm) and
  deduced values of $({v}/{c})^2$ for the different systems with flavour
  number $n_f$. }
\begin{center}
\begin{tabular}{ccccc|ccc}
$n_f$ & System  & $\kappa$ & $b$ & $\alpha$ & mass & $<r^2_{w}>^{1/2}$ 
& $({v}/{c})^2$ \\ 
\hline
1 &$\nu_e$   & 1.4 & 7.96 10$^{-9}$ & 2.0 & ~0.015 eV & 6.8 10$^{-9}$~ & 6.7
10$^{-38}$ \\  
2 &$\nu_\mu$  & 1.4 &  3.96 10$^{-9}$ & 2.0 & 0.01735 eV & 3.4 10$^{-9}$ & 2.2
10$^{-38}$  \\  
3 &$\nu_\tau$ & 1.4 & 2.25 10$^{-9}$ & 2.0 & 0.0512 eV & 1.9 10$^{-9}$ & 6.4
10$^{-38}$  \\ 
\hline
1 &$e$   & 1.4 &  3.95 10$^{-10}$ & 2.0 & ~0.51 MeV & 3.4 10$^{-10}$~ & 1.9
10$^{-25}$  \\  
2 &$\mu$  & 1.4 &  2.65 10$^{-10}$ & 2.0 & 105.7 MeV & 2.3 10$^{-10}$ & 3.7
10$^{-21}$  \\  
3 &$\tau$ & 1.4 & 1.98 10$^{-10}$ & 2.0 & 1777 MeV & 1.7 10$^{-10}$ & 5.9
10$^{-19}$  \\  
\end{tabular}
\end{center}
\end{table}

Table~1 shows indeed very small radii of the order of $10^{-9}$ fm. The
extracted value of $\alpha$$\sim$2 may be used to show a qualitative relation
between magnetic and electric bound states. For
electric binding of $q\bar q$ systems rather similar values of $\alpha$ of 1.5
and 2.4 have been extracted for $\Phi$ and $J/\psi$ mesons with a masses of
1.02 and 3.10 GeV, respectively, and mean radii square of 0.15 and 0.016
fm$^2$, see ref.~\cite{Mo2}. This  
may indicate that for $\nu_e$ the magnetic force gives rise to a bound state
energy smaller than the electric force by a factor of about 10$^{-11}$
and a radius smaller by a factor of 3 10$^{-8}$. 

The value of $({v}/{c})^2$ of several $10^{-38}$ leads to a total coupling
constant $\alpha_{tot}=\alpha^3 ({v}/{c})^2$ of the order of 10$^{-37}$, which
is only two orders of magnitude larger than the gravitational coupling $G_N
m_1m_2/\hbar c=6.707~10^{-39}$. For composite systems larger radii are
expected and the coupling constant should decrease further. This supports
the conclusion of a previous less constrained neutrino 
analysis~\cite{Moneu} that gravitation may be due to magnetic interactions 
of particles in matter. For a real proof of this conjecture calculations 
have to show that indeed the strict boundary conditions for magnetic bound
states are fulfilled for gravitational systems. Preliminary
studies~\cite{Mo1} show that rotational velocities of galaxies are
well described in the present approach (without dark matter contributions). 
However, it should be verified also that the complex dynamics of the present
Lagrangian with a mixing of motion and bound state creation is consistent
with astrophysical observations.   

For charged leptons a quite similar structure as for neutrinos is expected.
Again, for a pure $q^o\bar q^o q^-$ configuration no
interactions take place. If the $q^o\bar q^o$ pair decays to two $q^+ q^-$
pairs with opposite velocity v of positive and negative fermions, again a
magnetic bound state should be formed, for with the
conditions~(\ref{eq:mag1})-(\ref{eq:mag3}) are fulfilled. This is indeed the
case, indicating that also charged leptons can be regarded as 
magnetic bound states, with radii still smaller than of neutrinos. The results
are also given in table~1.

For the electron the boson-density and the potentials are given in the upper
part of fig.~3, which show that the boundary condition~(\ref{eq:conr}) is
well fulfilled. Also the Fourier transformed quantities are in good agreement,
as shown in the middle part. The shape of the confinement potential
$V_{2g}(r)$, displayed in the lower part, is also rather similar as for
neutrinos. Importantly, as seen in the upper part of fig.~3, the average
radius of several $10^{-10}$ fm is in agreement with results from
M$\phi$ller scattering~\cite{Moller}, from which the electron radis has been
estimated to be $<10^{-9}$~fm. 

Further, the magnetic bound state interpretation is in agreement with the
anomalous moments of charged leptons. Writing the magnetic moment by 
$\vec\mu_l=\frac{e\hbar}{2m_l} \sum_i \int r^2 dr \bar\psi_i(r)\vec 
\frac{s}{\hbar} \psi_i(r)$, the fact that two fermion densities are involved
yields $\vec\mu_l=2~\frac{e\hbar}{2m_l}$. This is consistent with the $e$ and
$\mu$ data, apart from a 1 $^o/_{oo}$ effect due to higher order mass
corrections~\cite{Schw}.      

\begin{table}
\caption{Comparison of mass ratio squares $F_m^2$, mean radius squares
  $F_{rad}^2$ and resulting values of $({v}/{c})^2$ for charged leptons in
  comparison with the results in table~1 for the three different
  charged/neutral lepton pairs.}
\begin{center}
\begin{tabular}{ccccc||c}
$n_f$ & System  & $F_m^2$ & $F_{rad}^2$ & $F_m^2 F_{rad}^2
({v}/{c})_\nu^2$ & $({v}/{c})_c^2$ \\  
\hline
1 & e, $\nu_e$     & 1.16 10$^{15}$ & 2.5 10$^{-3}$ & 1.9 10$^{-25}$ & 1.9
10$^{-25}$ \\ 
2 &$\mu, \nu_\mu$   & 3.71 10$^{19}$ & 4.6 10$^{-3}$ & 3.7 10$^{-21}$ & 3.7
10$^{-21}$ \\  
3 &$\tau, \nu_\tau$ & 1.21 10$^{21}$ & 8.0 10$^{-3}$ & 6.2 10$^{-19}$ & 5.9
10$^{-19}$ \\    
\end{tabular}
\end{center}
\end{table}
It is interesting to compare the flavour dependence of electric and magnetic
bound states. Whereas for electric $q\bar q$ bound states $\alpha^3$ 
increases as a fuction of $n_f$, see ref.~\cite{Mo2}, for magnetic bound
states the product $\alpha^3({v}/{c})^2$ should change. The results in 
table~1 show that only $({v}/{c})^2$ changes, whereas $\alpha$
stays constant. 

This independence on $\alpha$ for magnetic bound states allows to describe
the different masses and corresponding $({v}/{c})^2$ values by a simple
mass-radius relation. First, the relative velocity square of a
charged lepton $({v}/{c})^2_c$ may be proportional to that of the corresponding
neutrino multiplied with the square of the ratio of their masses
$F_m^2=(m_c/m_\nu)^2$. This cannot be entirely correct, because both the
masses and radii of the systems change. Taking both mass
ratio and $F_{rad}^2= <r^2_c>$/$<r^2_\nu>$ into account yields   
\begin{equation}
({v}/{c})_c^2 \simeq F_m^2 F_{rad}^2 ({v}/{c})_\nu^2 \ .
\label{eq:mradv}
\end{equation}
In table~2 the different ratios and products are given for the three flavour
charged/neutral lepton pairs. This shows that the differences of the
extracted velocities of charged and neutral leptons in table~1 are 
well described by the simple relation~(\ref{eq:mradv}). By this, the
different mass dependences of charged and neutral leptons in
fig.~1 are well understood, arising from orthogonality of the wave
functions of the different flavour states. 
Also it should be stressed that the good agreement between charged and neutral
leptons has been achieved by using the neutrino masses from the
ref.~\cite{Moneu}. This supports the correctness of the extracted neutrino
masses. 

Finally it is important to mention that due to the $(q\bar q)^nq$ structure
of leptons mesons of $(q\bar q)^n$ structure can decay only to
lepton-antilepton pairs, whereas baryons of $(q\bar q)^nq$ structure decay to 
baryon and lepton-antilepton pairs. These decays are weak due to an extremely
small overlap of the very different hadron and lepton wave functions. 
\vspace{0.6cm}

In conclusion, a new theoretical framework for the description of fundamental
forces has been discussed, based on a generalisation of electromagnetic
interactions. Compared to the SM, in which an understanding of the mass of
different flavour states of hadrons and leptons requires supersymmetric
fields, but also Higgs-field and Majorana neutrinos, in the present formalism
none of these fields are needed. 

With massless elementary fermions (quantons) and massless gauge bosons, a
direct coupling to the absolute vacuum with average energy $\tilde E_{vac}=0$
is obtained; further, severe boundary 
conditions for electric and magnetic bound states are fulfilled, by which all
parameters of the model are determined. This shows that all criteria of a
fundamental theory of relativistic particle bound states are fulfilled, which
leads most likely to a coherent description of all fundamental forces of
nature.    
For more extended studies it could be advantageous, if other solutions of the
Lagrangian~(\ref{eq:Lagra}) would be found, which go beyond the presently 
used evaluation of matrix elements.

Only a minimum of elementary fields is needed, one boson gauge field and
charged and neutral fermion (quanton) fields (of electric and magnetic
structure). Together with the fact that hadrons are understood as
electric and leptons by magnetic bound states, this emphasizes the inherent
symmetry of electic and magnet phenomena of Maxwell's theory also in
fundamental physics.

\begin{figure}
\centering
\includegraphics [height=18cm,angle=0] {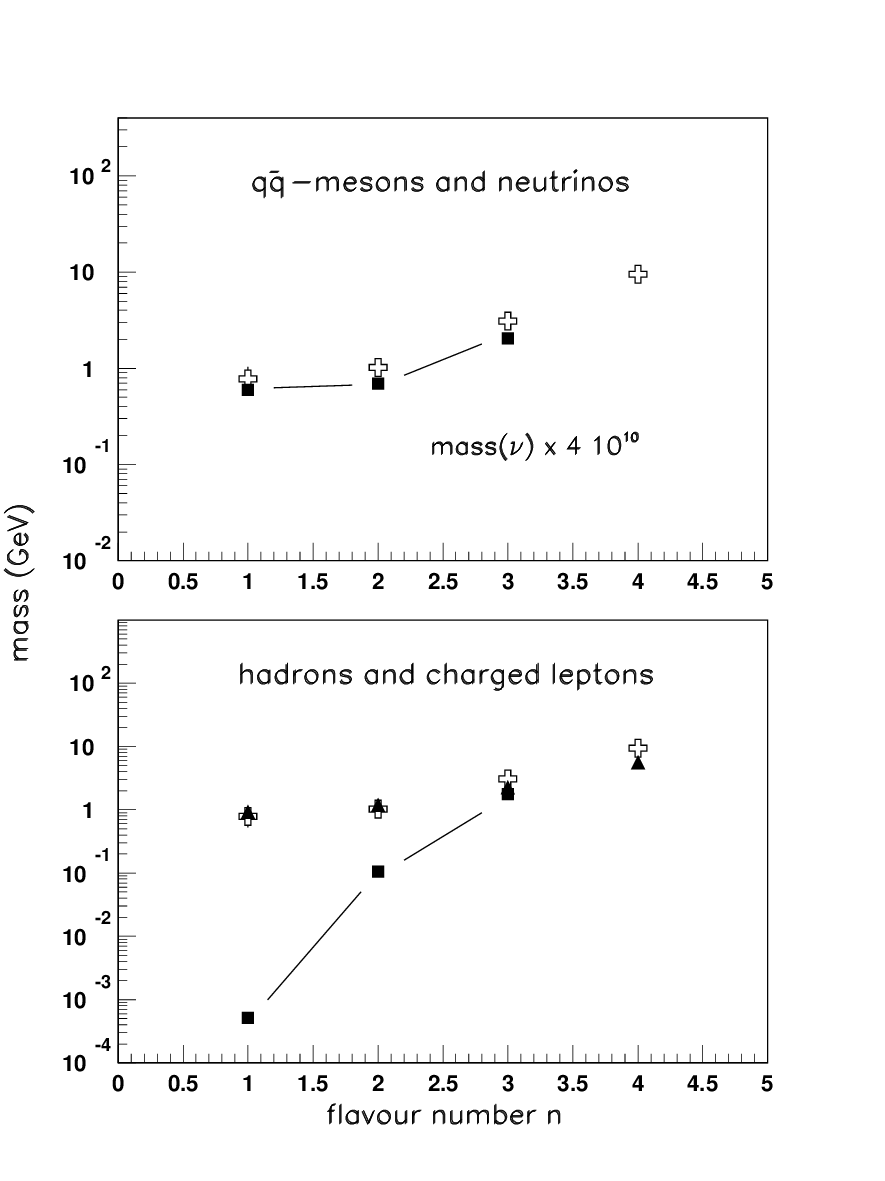}
\caption{Flavour dependence of hadron and lepton masses. \underline{Upper
    part:} $q\bar q$ meson masses (open crosses) as a function of the 
  flavour number $f_n$=1-4 ($\omega$, $\Phi$, $J/\psi$ and $\Upsilon$)
  and masses of neutral leptons taken from ref.~\cite{Moneu} ($f_n$=1-3 
  relate to $\nu_e$, $\nu_\mu$ and $\nu_\tau$), given by solid squares.
  \underline{Lower part:} Similar plot for baryons (triangles), $q\bar q$ 
  mesons (open crosses) and charged leptons (solid squares). 
  For baryons $f_n$=1-4 represent nucleon, $(\Lambda, \Sigma)$, $\Lambda_c$
  and $\Lambda_b$, for leptons $f_n$=1-3 correspond to $e$,
  $\mu$ and $\tau$. The straight lines are to guide the eye. }    
\label{fig:masshl}
\end{figure} 

\begin{figure}
\centering
\includegraphics [height=18cm,angle=0] {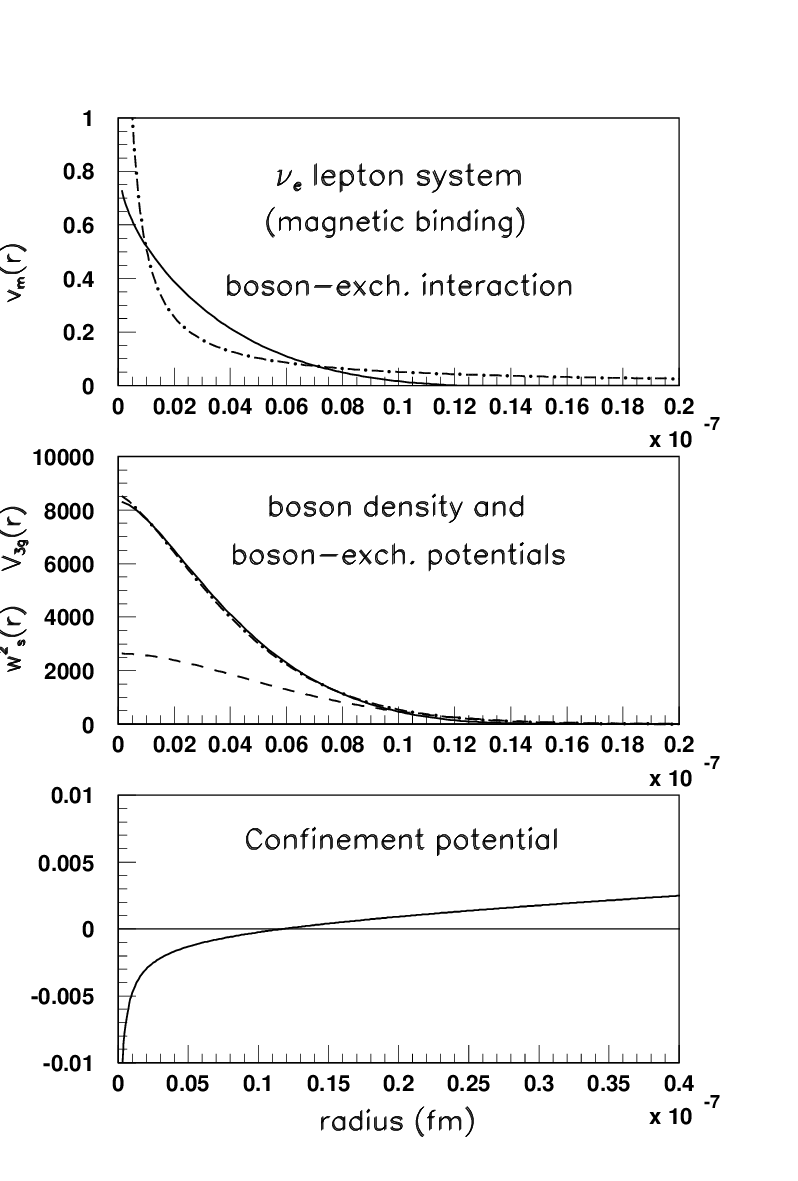}
\caption{Self-consistent solution for the $\nu_e$ system. 
  \underline{Upper part:} Relative interaction $v_v(r)$ in comparison with the
  Ampere potential $(\sim 1/r)$ given by dot-dashed line.
 \underline{Middle part:} Boson density $w_s^2(r)$ (dot-dashed line) and
 boson-exchange potentials $|V^{s,v}_{3g}(r)|$ given by dashed and solid
 lines, respectively. 
 \underline{Lower part:} Deduced confinement potential $V_{2g}(r)$. }    
\label{fig:g1exep}
\end{figure} 

\begin{figure}
\centering
\includegraphics [height=18cm,angle=0] {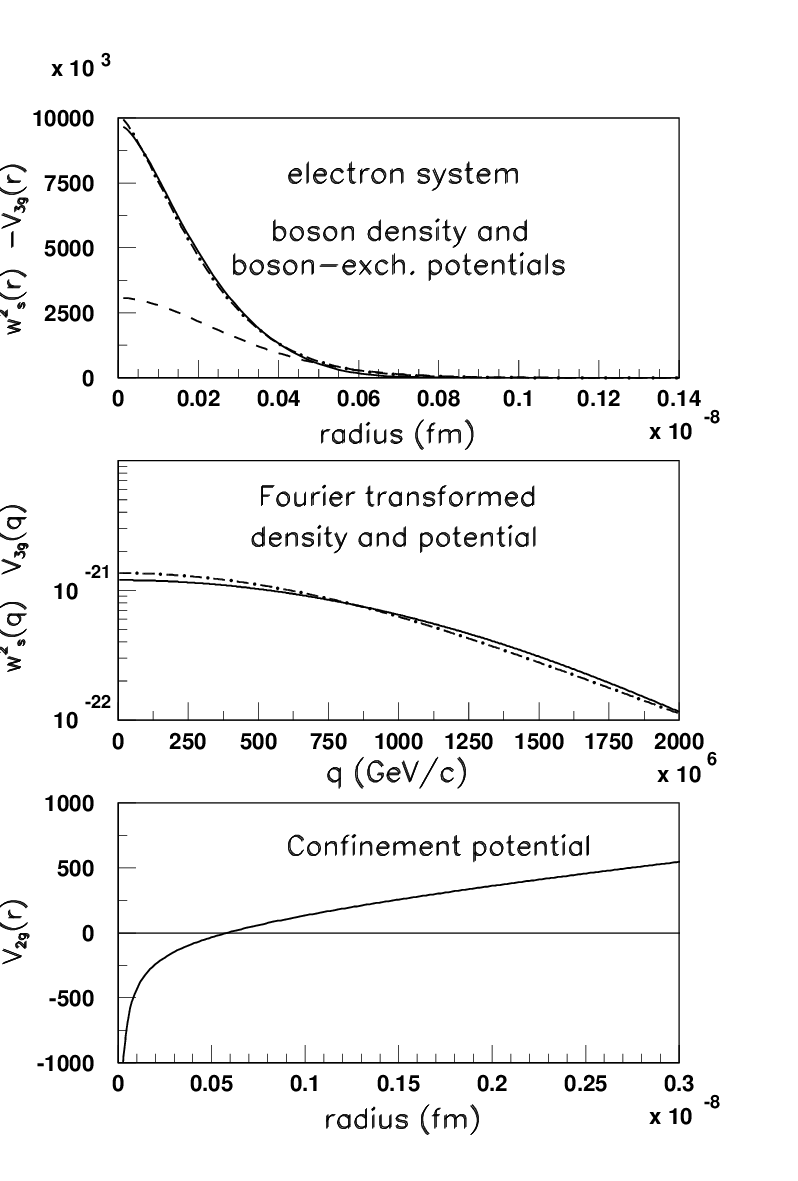}
\caption{Self-consistent solution for the electron system. 
  \underline{Upper part:} Boson density $w_s^2(r)$ (dot-dashed line) and
 boson-exchange potentials $|V^{s,v}_{3g}(r)|$ given by dashed and solid
 lines, respectively.
 \underline{Middle part:} Fourier transformed boson density $w_s^2(q)$
 (dot-dashed line) and boson-exchange potential $|V^{v}_{3g}(q)|$ given by 
 solid line. 
 \underline{Lower part:} Deduced confinement potential $V_{2g}(r)$. }    
\label{fig:g1exelec}
\end{figure} 

\end{document}